\def \SAIT #1 #2 {{\em Mem.\ Soc.\ Astron.\ It.\/} {\bf #1}, #2}
\def \MESS #1 #2 {{\em The Messenger\/} {\bf #1}, #2}
\def \ASTRNACH #1 #2 {{\em Astron. Nach.\/} {\bf #1}, #2}
\def \AAP #1 #2 {{\em Astron. Astrophys.\/} {\bf #1}, #2}
\def \AAL #1 #2 {{\em Astron. Astrophys. Lett.\/} {\bf #1}, L#2}
\def \AAR #1 #2 {{\em Astron. Astrophys. Rev.\/} {\bf #1}, #2}
\def \AAS #1 #2 {{\em Astron. Astrophys. Suppl. Ser.\/} {\bf #1}, #2}
\def \AJ #1 #2 {{\em Astron. J.\/} {\bf #1}, #2}
\def \ANNREV #1 #2 {{\em Ann. Rev. Astron. Astrophys.\/} {\bf #1}, #2}
\def \APJ #1 #2 {{\em Astrophys. J.\/} {\bf #1}, #2}
\def \APJL #1 #2 {{\em Astrophys. J. Lett.\/} {\bf #1}, L#2}
\def \APJS #1 #2 {{\em Astrophys. J. Suppl.\/} {\bf #1}, #2}
\def \APSS #1 #2 {{\em Astrophys. Space Sci.\/} {\bf #1}, #2}
\def \ASR #1 #2 {{\em Adv. Space Res.\/} {\bf #1}, #2}
\def \BAIC #1 #2 {{\em Bull. Astron. Inst. Czechosl.\/} {\bf #1}, #2}
\def \JSQRT #1 #2 {{\em J. Quant. Spectrosc. Radiat. Transfer\/} {\bf #1}, #2}
\def \MN #1 #2 {{\em Mon. Not. R. Astr. Soc.\/} {\bf #1}, #2}
\def \MEM #1 #2 {{\em Mem. R. Astr. Soc.\/} {\bf #1}, #2}
\def \PLR #1 #2 {{\em Phys. Lett. Rev.\/} {\bf #1}, #2}
\def \PASJ #1 #2 {{\em Publ. Astron. Soc. Japan\/} {\bf #1}, #2}
\def \PASP #1 #2 {{\em Publ. Astr. Soc. Pacific\/} {\bf #1}, #2}
\def \NAT #1 #2 {{\em Nature\/} {\bf #1}, #2}

\documentstyle{memsait}
\input epsf.sty
\begin{opening}
\title{THE TRICKY GALACTIC CLUSTER NGC2420} % ALL CAPITAL LETTERS PLEASE !!!
\author{P.G. Prada Moroni$^1$, V. Castellani$^2$, S. Degl'Innocenti$^2$, 
M. Marconi$^3$}
\institute{$^1$Dipartimento di Fisica, Universit\'a di Genova, 
and  INFN, Sezione di Genova, Genova, Italy\\
$^2$Dipartimento di Fisica, Universit\'a di Pisa, 
and  INFN, Sezione di Pisa, Pisa, Italy\\
$^3$Osservatorio Astronomico di Capodimonte, Napoli, Italy}
\date{} % DO NOT INSERT ANY DATE HERE !!!
\end{opening}

\begin{document}

%\oddpagefooter{\sf Mem. S.A.It., Vol. ??, ??}{}{\thepage}
%\evenpagefooter{\thepage}{}{\sf Mem. S.A.It., Vol. ??, ??}
\oddpagefooter{}{}{} % LEAVE AS IT IS !
\evenpagefooter{}{}{} % LEAVE AS IT IS !
\ 
\bigskip

\begin{abstract}
We discuss the CM diagram of the galactic cluster NGC2420 to the light
of current theoretical predictions. By relying on the most recent
updating of the physical input, one finds too luminous theoretical He
burning stars together with the evidence for a misfitting of the lower
portion of the MS. Moreover one finds two well known overshooting
signatures, as given by i) the large extension of the ``hook" preceding
the overall contraction gap, and ii) the scarcity of stars just at the
end of the gap. We show that the overluminosity of He burning stars
appears as a constant prediction of models based on updated physics,
whereas alternative assumptions about the Equation of State can
account for the MS fitting. Moreover, due to the scarse statistical
significance of the observational sample, one finds that overshooting
signatures can be present also in canonical (without overshooting)
predictions. We conclude that, unfortunately, NGC2420 does not keep
the promise to be of help in costraining the actual dimensions of
convective cores in H burning MS stars, suggesting in the meantime
that using clumping He burning stars as theoretical standard candle is
at least a risky procedure.  In this context the need for firmer
constraints about the reddening of galactic clusters is shortly
discussed.
\end{abstract}

\section{Introduction}
The beautiful CM diagram presented in 1990 by Anthony-Twarog et al.
for the intermediate age open cluster NGC2420 has been in the
last ten years a favourite target for all the people concerned with
the evolution of low to intermediate mass stars. 
The occasion for revisiting this cluster has been
given to us by the recent paper by Pols et al. (1998, hereinafter P98), who presented
new evolutionary tracks carefully discussing the fit of a selected (and
well chosen) sample of galactic clusters. As discussed in that paper,
NGC2420 seems to give a good chance to put firm constraints on the
efficiency of the mechanism of core overshooting.
Owing to the relevance of such an issue, 
we decided to go deep into the matter, hoping eventually  to settle 
down such a long debated argument. However, as we will discuss 
in the following, the situation is far from being assessed.

\section{Isochrone fitting}

According to Anthony-Twarog et al. (1990) metallicity estimates give
for the cluster [Fe/H]=-0.35 $\pm$0.10, but with evaluations reaching
[Fe/H]=-0.6 (Canterna et al.  1986). Following P98, we will assume
Z=0.007 ([Fe/H=-0.4]).  By taking Y=0.23 for old metal poor stars and
accounting for a galactic He enrichment law as given by $\Delta
Y/\Delta Z \approx 2.4$ (see e.g. Pagel \& Portinari 1998) we obtain
for the cluster Y=0.244.  The adopted evolutionary code and physical
inputs are the same as in Cassisi et al. (1998); in particular to start
our investigation we adopted the OPAL EOS (Rogers et al. 1996).

\begin{figure}
\label{fit1}
\epsfysize=5cm 
\epsfbox{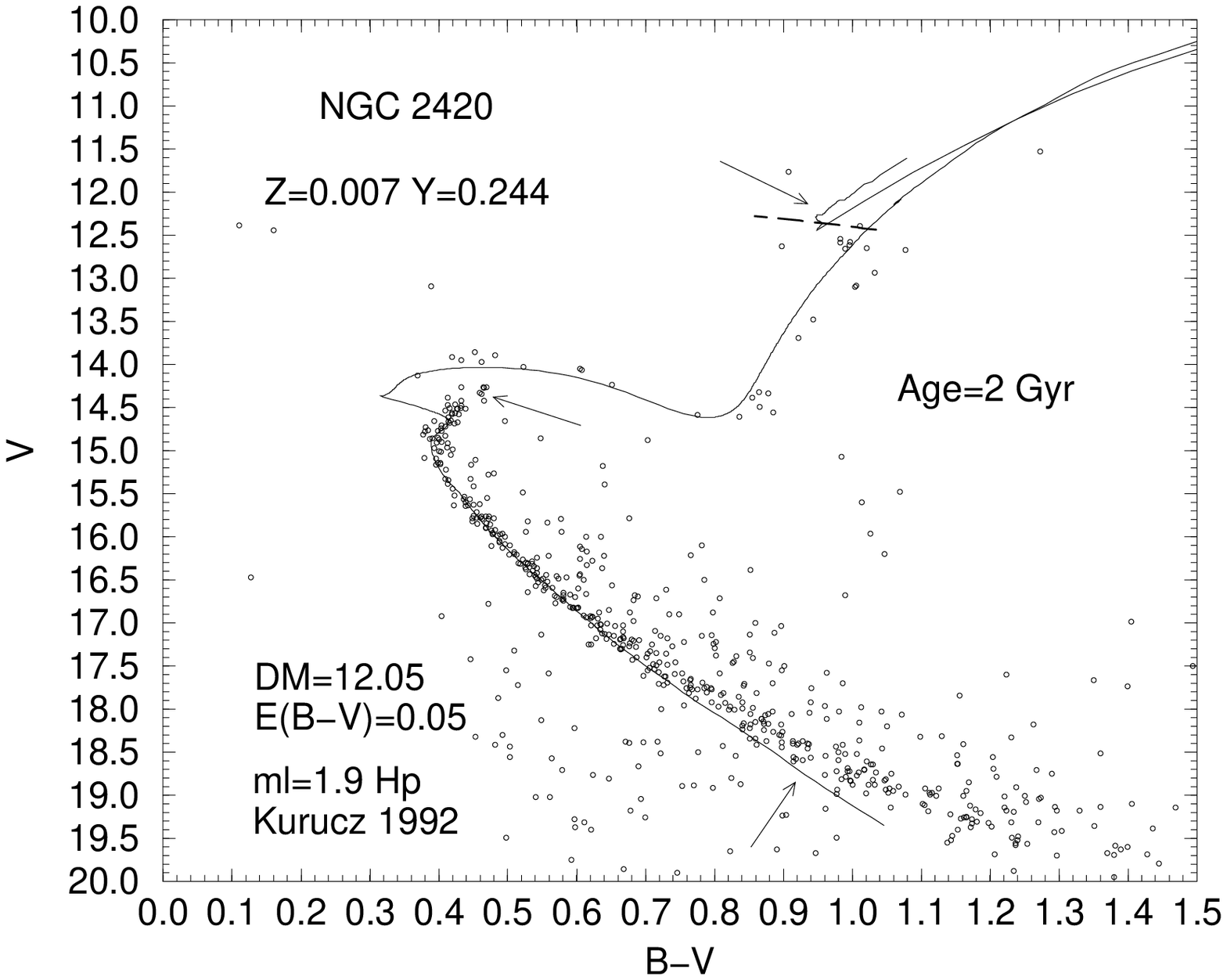}
\vspace{-5cm}
\hspace{7cm}
\epsfysize=5cm
\epsfbox{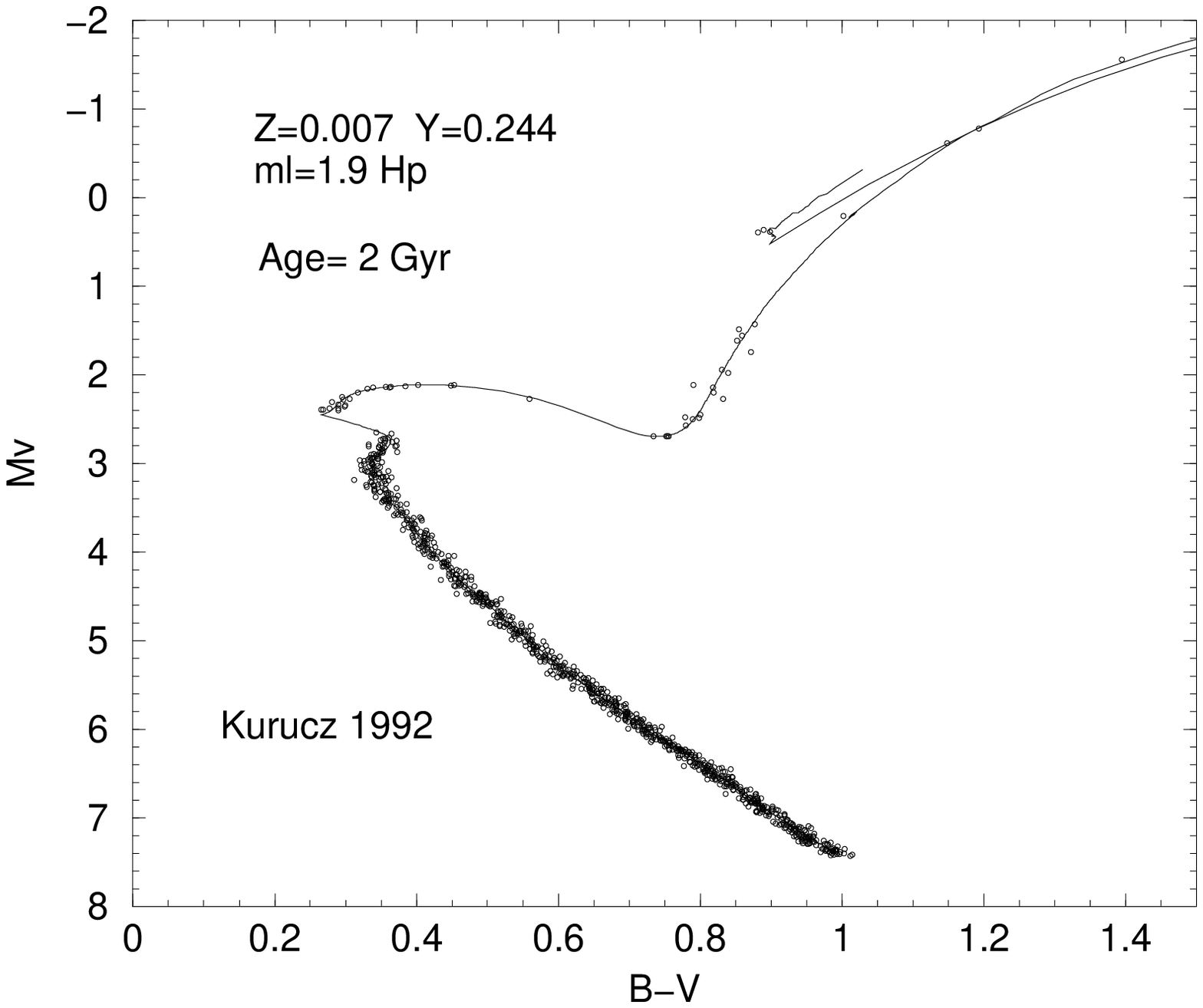}
\caption{Left panel:Fit of the cluster with the 2.0 Gyr
theoretical isochrone;
 the arrows 
mark the regions where the fitting is not good. The dashed line indicates
the variation of the visual magnitude of the He clump when its
temperature is varied to reach the observational color of He burning stars,
see text. Right panel: The same  2.0 Gyr isochrone
 with overimposed the corresponding
synthetic CM diagram without binaries.
}  
\end{figure}

Figure 1 (left panel) shows the best fitting of our isochrone to the
cluster CM diagram, as obtained by using colour-temperature relations
from Kurucz (1992) and by adopting for the cluster just the same
reddening and the same age as in P98.  As a result, we found that a
best fitting is indeed achieved, with only a small difference in the
cluster distance modulus (DM = 12.05 against 11.95 in P98), which
appears the natural result of the differences between the two MS
luminosities already discussed in Prada Moroni et al. (2000).
 The impressive agreement between the two theoretical scenarios
can obviously taken as an evidence that neither differences
in the adopted input physics nor in the color-temperature relation  
play a relevant role in the predicted models. As
already discussed by P98, one finds that theoretical He burning stars
appear too luminous, by about 0.3 mag. However, in both theoretical
isochrones one can also find the evidence for a misfitting of the
lower portion of the MS, with too blue theoretical models. As well
known, low masses MS stars are less and less affected by the efficiency
of the external convection, thus it appears difficult to reconcile
theory and observation only by tuning the mixing lenght parameter (see
e.g.  Castellani et al. 1999).  To discuss overshooting, Figure 1
(right panel) depicts theoretical expectations, showing a synthetic
cluster as obtained by randomly populating the isochrone till
obtaining 24 stars brighter than ${M}_{V}=2.1$, as
observed. Comparison with the CM diagram in Fig. 1 (left panel)
discloses in the observational data the two well known overshooting signatures,
namely: i) the larger extension of the ``hook'' preceding the overall
contraction gap, and ii) the lack of the concentration of stars just
at the end of the gap, expected under canonical assumptions.

\begin{figure}
\label{fit3}
\epsfysize=5cm 
\epsfbox{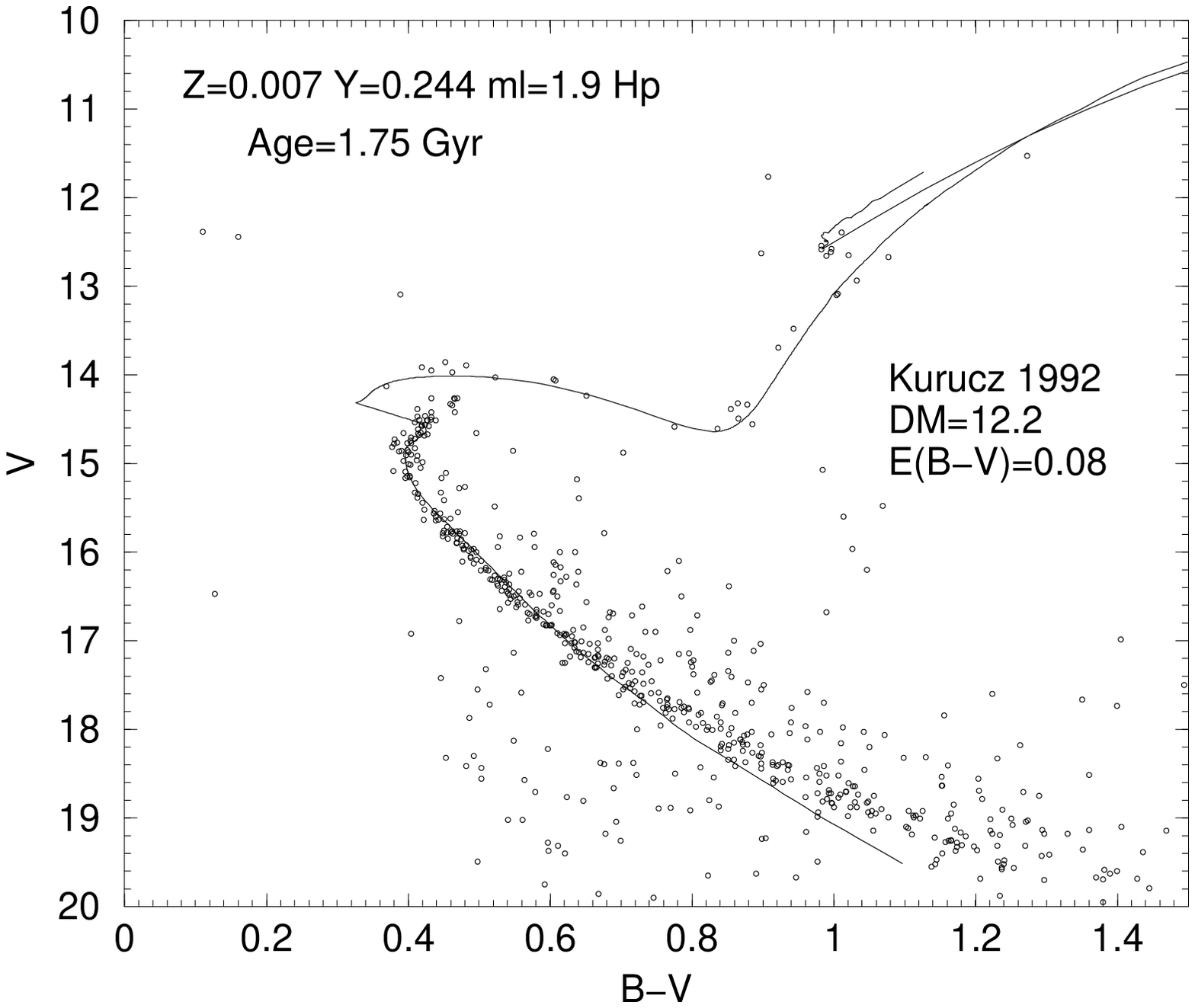}
\vspace{-5cm}
\hspace{7cm}
\epsfysize=5cm
\epsfbox{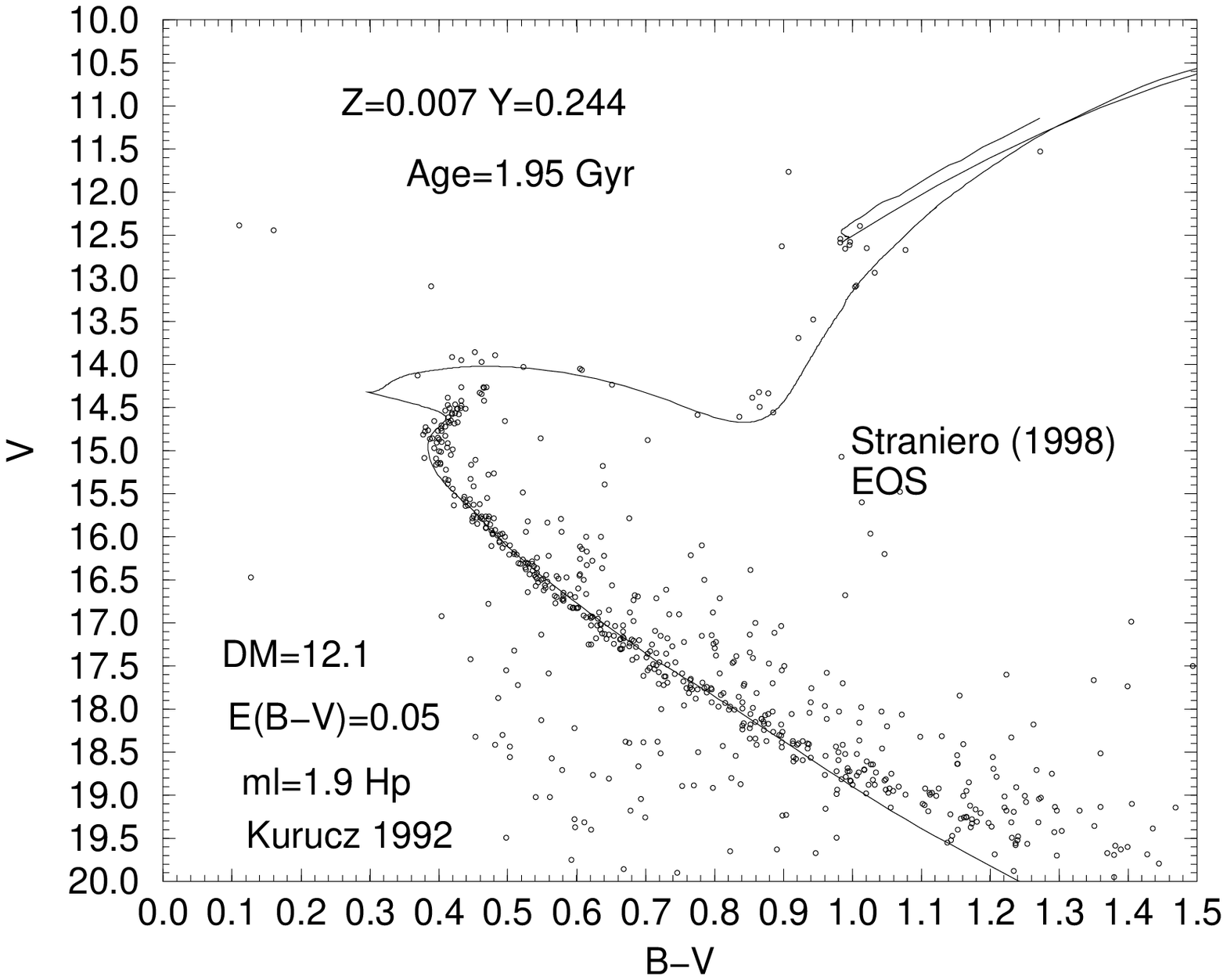}
\caption{Left panel: Best fit of the cluster with
1.75 Gyr theoretical isochrone with OPAL EOS.
 Right panel: Best fit of the cluster with 1.95 Gyr
theoretical isochrone with Straniero (1998) EOS} 
\end{figure}

However, the fitting can be improved by increasing the reddening thus
 driving a larger evaluation of the cluster distance modulus,
 alleviating the overluminosity of predicted He burning stars. This is
 shown in Fig. 2 (left panel), which reports our best fitting as
 obtained for E(B-V) =0.08. Moreover, increasing the reddening the
 Turn Off becomes bluer and the cluster younger, increasing the
 extension of the hook (see Fig. 1 for comparison), also alleviating this
 discrepancy. However, no one of these discrepancies is completely
 resolved, nor increasing the reddening helps to solve the
 disagreement for the lower MS.
Concerning the problem of the MS fitting, one has to conclude that either 
the color temperature relation or the MS models are wrong, if not
both. To explore theoretical models, we repeated the previous  
computations
but adopting the alternative Equation of State presented by 
Straniero (1998). As shown in Fig. 2 (right panel),
 we found that theoretical and observational MSs
appear now in reasonable agreement, disclosing that a solution to this
problem already exists within the currently available theoretical
scenarios.  On the contrary, the replacement of Kurucz (1992)
colour-temperature relations with the most recent and widely adopted
semiempirical relations by Alonso et al. (1996, 1999) cannot solve
the dicrepancy between the theoretical and the observational low MS.
In this case, in fact, one obtains a shift toward bluer MS
colours, which would affect (increase) the reddening evaluations by
about 0.05 mag, but not the MS slope.

\section {The tricky binaries}

According to the discussion in the previous section, one is left with
the evidence of a small overluminosity of He burning stars and with
two clear signatures for the efficiency of overshooting. As a matter
of the fact, numerical experiments confirmed that allowing for a
moderate overshooting (of the order of 0.1 pressure scale height, Hp)
our theoretical isochrone would match much better the observed
diagram.  However, in a previous paper (Castellani et al. 1999) we
have already drawn the attention on the evidence that binaries stars,
as obviously present in the cluster, may mimike overshooting by
extending the hook above its natural maximum luminosity. Thus we
tested the effects of binarity again by producing synthetic
clusters from canonical (without overshooting) models but allowing for 30\%
binaries (as obtained by a rough estimate from the CM diagram)
according to the following distribution of mass ratio: 50\% between
0.8 and 1, 30\% between 0.6 and 0.8, the remaining 20\% totally random
(Tornamb\`e A., private communication).

\begin{figure}
\label{fit1}
\epsfysize=5cm 
\epsfbox{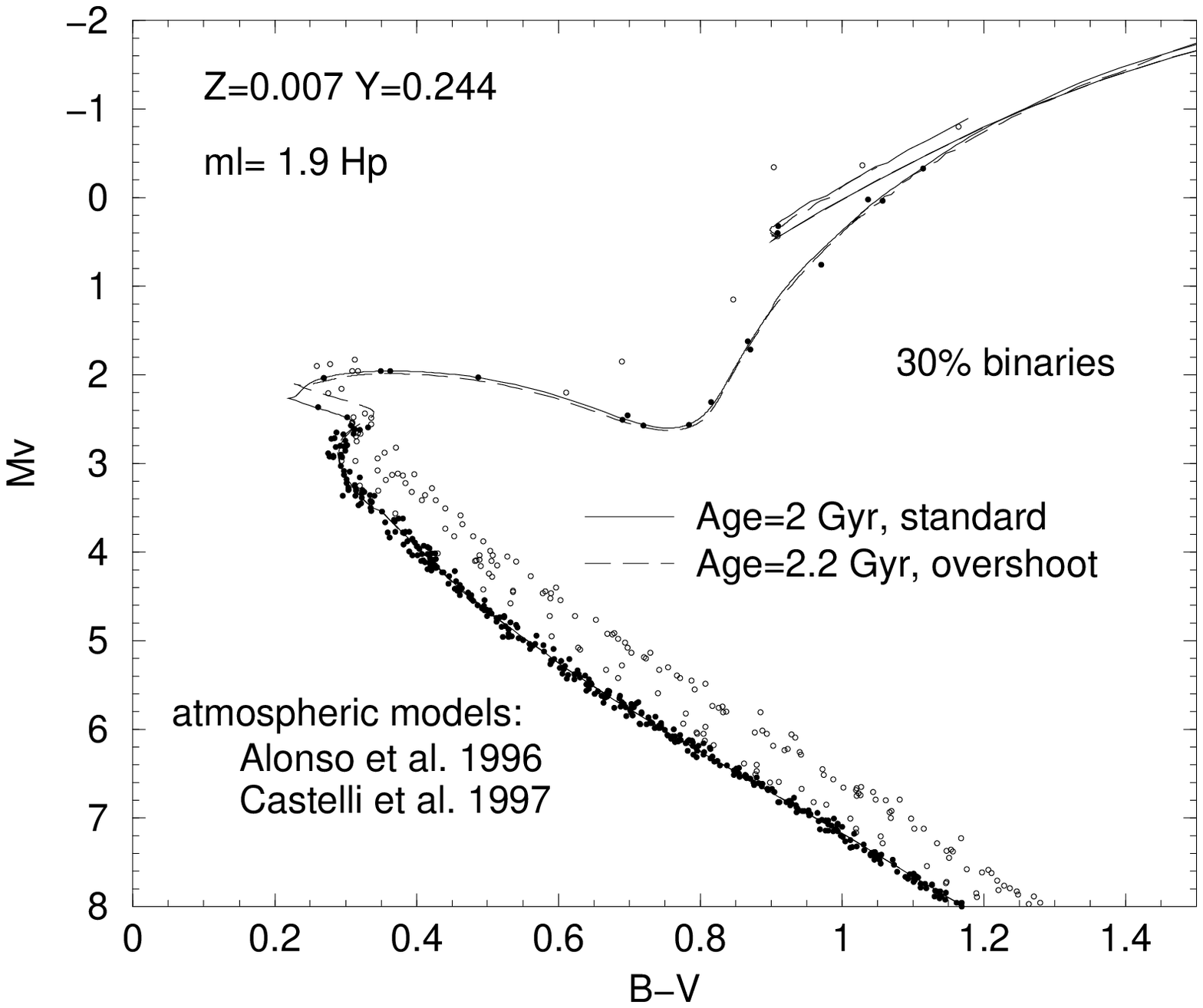}
\vspace{-5cm}
\hspace{7cm}
\epsfysize=5cm
\epsfbox{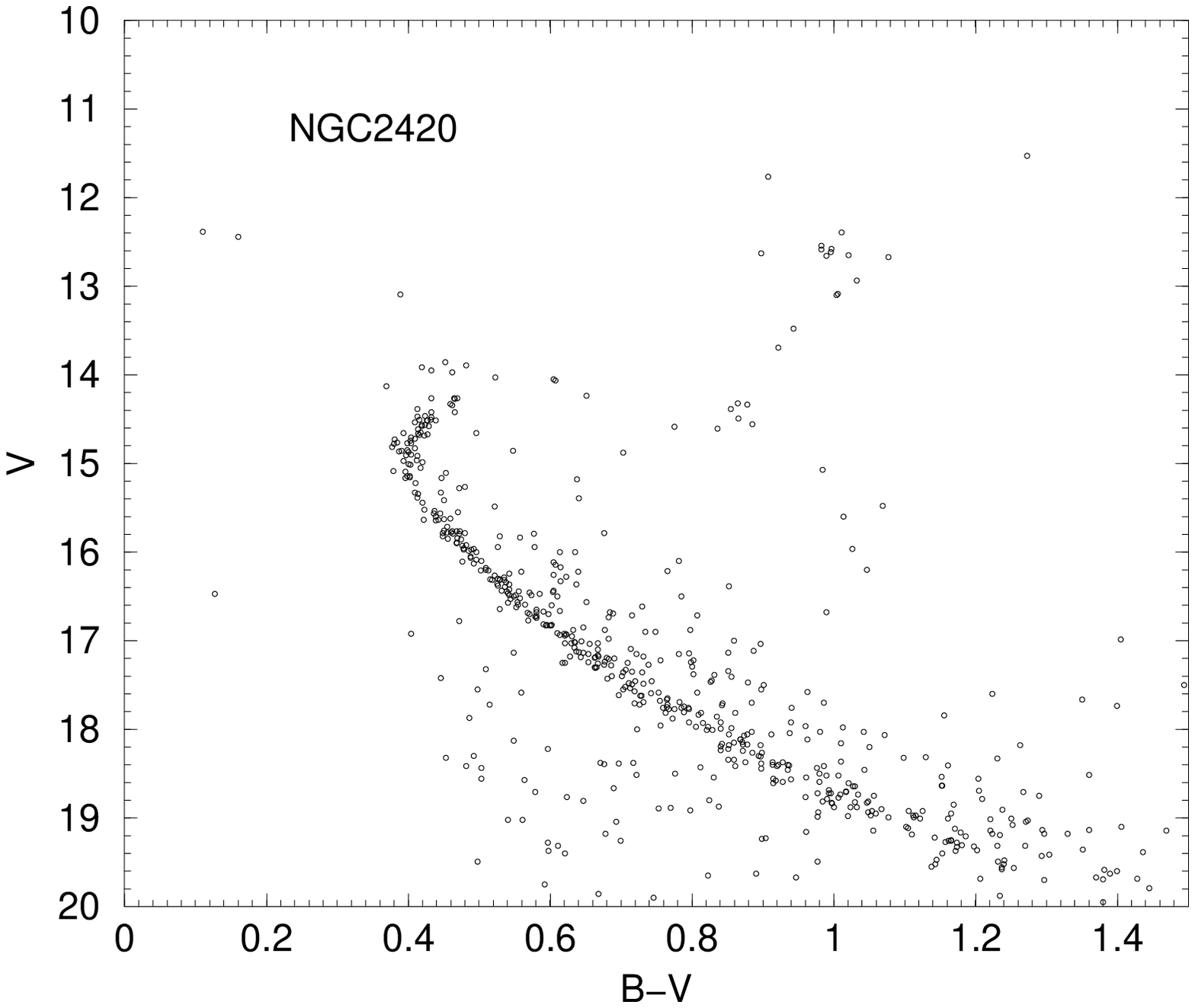}
\caption{Left panel: solid line: theoretical standard (without overshoot)
 2.0 Gyr isochrone with Straniero (1998) EOS (Z=0.007 Y=0.244) with
 overimposed the corresponding synthetic CM diagram including 30\% of
 binaries; dashed line: theoretical 2.2 Gyr isochrone with Straniero
 (1998) EOS and mild overshooting (${l}_{ov}=0.1
 {H}_{p}$). Right
 panel: NGC2420 CM diagram by Anthony-Twarog et al. 1990.
 Observational data are not dereddened.}
\end{figure}
As predicted, and as shown in Fig. 3, left panel, binaries 
tend to extend the hook, producing in this region a CM diagram which appears 
in better agreement with overshooting computations. However,
and to our surprise, one also finds that in a good percentage of
the random clusters, namely of the order of about 50\%, the predicted 
``post-gap" distribution sensitively deviates from canonical expectations,  
approaching the one actually observed in the cluster (see, e.g., 
Fig. 3, right panel). The reason for such an occurrence is easily 
found in the scarce statistical significance of the number of 
luminous stars together with the non negligible contribution of
binaries. As a result, we conclude that NGC2420 is actually failing 
to give the expected constraints about the overshooting efficiency.
As a matter of the fact, both the discussed canonical and 
mild overshooting cases appear acceptable.
However, numerical simulations seems to suggest that an overshooting
as large as 0.25 Hp would depopulate too much the red part of the subgiant branch.

\section{Conclusions}
According to the previous sections, one finds that three out of the
four theoretical misfitting of the cluster CM diagram can be
accounted for within current evolutiony scenarios.
On the contrary, no assumption appears able to reconcile the predicted
luminosity of He burning stars with observation.
To explore all the possibilities one may guess that a given amount of
mass loss could account for this discrepancy. One generally assumes
that mass loss occurs in the advanced phase of H shell burning, so
that the internal structure of the He burning star is not affected by
such an occurrence, which only decreases the amount of envelope
surrounding the central He core. Under this assumption, the effect of
mass loss on He burning models can be easily computed by simply
decreasing the envelope of the constant-mass model. Numerical
simulations shows that to reach the agreement between theory and
observation for the clump luminosity one needs to decrease the He
burning mass from the standard value of $\approx$ 1.5 M$_{\odot}$ to
1.0 M$_{\odot}$.  This very high amount of mass loss seems very
unlikely to us.

At present time the discrepancy between theory and observation in the
luminosity of He burning stars with degenerate RGB progenitors 
it is not a surprising result; it appears as a constant prediction of
 models based on updated physics (see e.g. P98, Castellani et al. 2000), whereas similar
suggestions for the overluminosity of theoretical models have been
also derived from the pulsational properties of RR Lyrae 
(see e.g. Caputo et al. 2000). We conclude that using clumping He
burning stars as theoretical standard candles, as we did (Castellani
et al. 1999), is at least a risky procedure. We note that in the above
quoted work, one could recognize a signature of the overluminosity in
the need of assuming rather large reddenings for all the clusters. In
this context, it follows that firm constraints about the reddening of
galactic clusters will help in solving this problem.

\end{document}